\documentclass
[superscriptaddress,secnumarabic,amssymb,amsmath,nobibnotes,aps,prd,showkeys,nofootinbib,nopacsnumber,onecolumn,12pt]{revtex4}%
\usepackage{graphicx}
\usepackage{bm}
\usepackage{amsmath}
\usepackage{amsfonts}
\usepackage{amssymb}%
\providecommand{\U}[1]{\protect\rule{.1in}{.1in}}
\newcommand{\be}{\begin{equation}}
\newcommand{\ee}{\end{equation}}

\newcommand{\mincir}{\raise
-3.truept\hbox{\rlap{\hbox{$\sim$}}\raise4.truept\hbox{$<$}\ }}
\newcommand{\magcir}{\raise
-3.truept\hbox{\rlap{\hbox{$\sim$}}\raise4.truept\hbox{$>$}\ }}

\begin{document}
\title{ Reconstruction of $\Lambda$CDM Universe from Noether symmetries in Chameleon gravity}
\author{Andronikos Paliathanasis}
\email{anpaliat@phys.uoa.gr}
\affiliation{Institute of Systems Science, Durban University of Technology, Durban 4000,
South Africa}
\affiliation{Departamento de Matem\'{a}ticas, Universidad Cat\'{o}lica del Norte, Avda.
Angamos 0610, Casilla 1280 Antofagasta, Chile}

\begin{abstract}
We apply the Noether symmetries to constrain the unknown functions of
chameleon gravity in the cosmological scenario of a spatially flat
Friedmann--Lema\^{\i}tre--Robertson--Walker space-time with an ideal gas. For
this gravitational model the field equations admit a point-like Lagrangian
with as unknown functions the scalar field potential and the coupling function
which is responsible for the chameleon mechanism. Noether's first theorem
provides us with four sets of closed-form functional forms for which
variational symmetries exist. We construct the corresponding conservation laws
and we use them in order to determine new analytic solutions in chameleon
gravity. From the analysis of the physical properties of the new solution it
follows that in the late universe they can reproduce the $\Lambda$CDM model
without having to assume the presence of a pressureless fluid in the
cosmological fluid.

\end{abstract}
\date{\today}
\maketitle

\section{Introduction}

\label{sec1}

The physical origin of the dark energy mechanism which is responsible for the
late-time acceleration phase of the universe puzzles the scientific community
\cite{q1,q2}. Cosmologists have proposed various theoretical models to explain
dark energy \cite{q3,q4}. Gravitational theories with a scalar field have
special relevance because they provide a very simple mechanism for the
description of the cosmological observations at the late universe
\cite{q5,q6,q7}. Within, scalar fields have been used for the description of
inflation through the inflaton mechanism \cite{q8,q9}. The novelty of scalar
fields is that they provide a dynamical dark energy scenario which can surpass
the major problems of $\Lambda$-cosmology \cite{q10}. The degrees of freedom
introduced in the field equations from the scalar fields can provide the
dynamical variables related to the modification of the Einstein-Hilbert Action
Integral \cite{q11,q12,q13}.

Quintessence is a very simple dark energy candidate described by a minimally
coupled canonical scalar field with a nonzero potential function \cite{q14}.
In quintessence cosmology acceleration occurs when the scalar field varies at
a slow rate. There is a plethora of quintessence models which have been
proposed in the literature for different potential functions which give a
different cosmological history and evolution \cite{q15,q16,q17}. The
quintessence model can describe the Chaplygin gas (-like) fluids
\cite{q18,q19,q20}, as to provide matter components in the cosmological fluid
\cite{q21}. There are various extensions of the quintessence model in the
literature. The phantom scalar field has a similar Lagrangian to the
quintessence model, but it has a negative kinetic energy term. The latter
leads to cosmological behaviour which a faster expansion of the universe is
provided by the exponential expansion of the de Sitter solution, as a result
of which this fast expansion is related to the Big\ Rip singularity
\cite{q22,q23}. For other scalar field models we refer the reader to
\cite{q24,q25,q26,q27,q28,q29,q30} and references therein.

In \cite{ch1,ch2} an extension has been proposed of the quintessence scalar
field model with a chameleon mechanism for the mass of the scalar field.
Specifically, in the cosmological scenario of General Relativity with a scalar
field and an ideal gas, a coupling function which depends upon the scalar
field as been introduced at the Lagrangian of the ideal gas. As a consequence
the mass of the scalar field depends upon an effective potential function with
the contribution of the coupling function and the energy density of the
ideal.\ This mechanism gives to the scalar field a small mass in open space
with low energy density and a large mass on compact objects such on the
surface of the Earth. Because of that property, the magnitude for the
violation of the equivalence principle and of the fifth force components are
different in the open space and in the solar system which provides a
\textquotedblleft hide\textquotedblright\ mechanism to the scalar field. That
is why it is called \textquotedblleft chameleon\textquotedblright.

The coupling function responsible for the chameleon mechanism in \cite{ch1}
has been considered to be of the exponential form. However, this chameleon
model \cite{ch1} has its origin on a more geometrical background.
Specifically, the chameleon model with the exponential coupling is equivalent
to the Weyl integrable spacetime (WIS) \cite{salim96}. In WIS the scalar field
is introduced in the gravitational Action Integral with a geometric
construction method. The scalar field in WIS\ plays the role of the conformal
factor between two conformally related metrics. The gravitational Lagrangian
is constructed by the Levi-Civita connection of the conformal metric which
provides the scalar field degrees of freedom in the field equations. In the
case that there is no matter source the WIS\ reduces to General Relativity
with a scalar field. There exists a difference between the chameleon mechanism
\cite{ch1} and the WIS \cite{salim96}. In the chameleon mechanism the scalar
field is a quintessence with positive kinetic density, on the other hand, in
WIS the scalar field can have negative kinetic energy, because a new parameter
has been introduced in the kinetic term of the scalar field Lagrangian.

The chameleon mechanism has been applied in cosmology as a dark energy
alternative. The mechanism provides an interaction between the elements of the
dark sector of the universe. Coupled dark energy models have drawn the
attention of cosmologists because they can gives a theoretical description on
the analysis of the cosmological observations such as the cosmic coincidence
problem \cite{ame1,ame2,pavon,delcampo}, gravitational waves observations
\cite{gwo} or the $H_{0}$ tension \cite{hot,hot2,hot3,hot4}. In
\cite{Wetterich-ide1} it was found that the interaction is a mechanism which
can lead to a dynamical cosmological constant term. There is a plethora of
studies in the literature which deal with the WIS or the chameleon scalar
field on cosmological scales. In \cite{w1} it has been shown that the
chameleon scalar field can play the role of the inflaton field related to the
inflationary era. This study shows that inflation has a geometric origin
related to the conformally related metric. The conditions for the late-time
acceleration in WIS were determined \cite{w2}; nevertheless in \cite{w3} it
has been derived that the scalar field in chameleon theory can have a phantom
behaviour, which lead to a Big Rip singularity \cite{w4}. The chameleon
mechanism provides the limit of hyperbolic inflation when the matter source is
described by a stiff fluid, that is, by a massless scalar field \cite{hy}. In
this limit the chameleon mechanism is reduced to the Chiral model
\cite{hy1,hy2}. The coupling function between the scalar field and the matter
source can take other functional forms beyond the exponential coupling
\cite{w5,w6,w7}. Over and above the chameleon mechanism it has been introduced
in other scalar field theories such as in k-essence, in tachyonic fields and
others, see for instance \cite{w8,w9,w10} and references therein. Before, we
mentioned that the chameleon mechanism contributes to the mass of the scalar
field in compact objects; however, in cosmological scales when the dark matter
(or the radiation) dominates to the universe during the chameleon mechanism
has a non-trivial contribution, which means that it can affect the dynamics in
large scales. A direct observation of the chameleon mechanism is presented in
\cite{dch1}

Although the field equations for chameleon gravity in a
Friedmann--Lema\^{\i}tre--Robertson--Walker universe are of second-order,
there are very few known exact and analytic solutions in the literature.
Recently in \cite{ns1} analytic solutions were determined for the case of WIS
with the use of the variational symmetries, while the method of variational
symmetries has been applied also and in the limit of the Chiral model
\cite{ns2}. Specifically the variational symmetry conditions are applied in
order to constrain the unknown functions and parameters of the gravitational
Lagrangian. Such a method has been widely used in cosmology with many
interesting results \cite{ns2,ns3,ns4,ns5,ns6,ns7,ns8}. For a detailed review
of the application of the variational symmetries in gravitational physics we
refer the reader in \cite{ns9}.

In this work we apply the variational symmetries, aka Noether symmetry
analysis, for the geometric constraint of the scalar field potential and of
the coupling function in chameleon cosmology. We perform a classification
problem as established by Ovsiannikov in \cite{ovs}. A recent attempt to
answer this classification problem is presented in \cite{ns10}. However, the
authors considered the ideal gas important for the chameleon mechanism to be
the cosmological constant. In this case the chameleon mechanism is cancelled
and the theory is equivalent to the quintessence model \cite{ns11}. The field
equations in chameleon cosmology are of second-order and they form an
autonomous system which admits a point-like Lagrangian function.\ According to
Noether's second theorem, for every variational symmetry there exists a
corresponding conservation law. The latter can be used to investigate the
integrability properties of the field equations and to determine similarity
transformations which are applied to reduce the order of the dynamical system.
The variational symmetries in this study lead to new analytic and exact
solutions in chameleon cosmology. The derivation of analytic solutions in
closed-form expression is important because we can study the behaviour of the
given gravitational theory by using analytic techniques. The structure of the
paper follows.

In Section \ref{sec2} we present the gravitational model of our consideration
which is that of a minimally coupled scalar field cosmology in a spatially
flat Friedmann--Lema\^{\i}tre--Robertson--Walker (FLRW) geometry with a
chameleon mechanism. The scalar field is coupled to the Lagrangian function of
the matter source. For the latter we assume that it describes an ideal gas.
For this model the field equations are of second-order and they are described
by a point-like Lagrangian with two unknown functions, the scalar field
potential and the coupling function which defines the chameleon mechanism. In
Section \ref{sec3} we apply the theory of variational symmetries, i.e. Noether
symmetries, as a selection rule to constrain the unknown functions of the
theory. We found that there exist four sets of functions for which the field
equations admit nontrivial Noether symmetries. For these cases we use
Noether's second theorem to write the admitted conservation laws. For the new
cosmological models derived by the symmetry analysis in Section \ref{sec4} we
derive the analytic solution for the field equations. We investigate the
evolution of the cosmological parameters and we show that these new integrable
models can explain the present acceleration phase of the universe, while these
chameleon models mimic the $\Lambda$CDM cosmology either if we have considered
the ideal gas to be different from that of a pressureless fluid related to the
cold dark matter. Finally, in\ Section \ref{sec5} we summarize our results.

\section{Chameleon cosmology}

\label{sec2}

Consider now a four-dimensional Riemannian manifold $V^{n}$ with metric tensor
$g_{\mu\nu}$ and Ricci scalar $R=R\left(  g\right)  $. Furthermore, we assume
the existence of a scalar field $\phi\left(  x^{\mu}\right)  $ and of a matter
source described by the Lagrangian function $L_{m}\left(  \psi\left(  x^{\mu
}\right)  \right)  $.

For the above, in the context of General Relativity we assume the
gravitational Action Integral \cite{w5}
\begin{equation}
S=\int\sqrt{-g}d^{4}x\left(  \frac{R}{2}-\frac{1}{2}g^{\mu\kappa}\nabla_{\mu
}\phi\left(  x^{\nu}\right)  \nabla_{\kappa}\phi\left(  x^{\nu}\right)
-V\left(  \phi\right)  -F\left(  \phi\right)  L_{m}\left(  x^{\nu}\right)
\right)  , \label{ac.01}%
\end{equation}
where $F\left(  \phi\right)  $ is the coupling function between the scalar
field and the matter source and describes the chameleon mechanism.

Variation with respect to the metric tensor provides the gravitational field
equations of General Relativity%
\begin{equation}
G_{\mu\nu}=T_{\mu\nu}^{eff},
\end{equation}
where $T_{\mu\nu}^{eff}$ is the effective energy-momentum tensor which
describe the degrees of freedom from the scalar field and the matter source.

Moreover, the equation of motion for the scalar field is%
\begin{equation}
\Delta\phi=V_{,\phi}+f_{,\phi}T_{\mu}^{\mu} \label{as.0}%
\end{equation}
in which $T^{\mu\nu}=\frac{\partial L_{m}}{\partial g_{\mu\nu}},$ and $\Delta$
is the Laplace operator for the background metic $g_{\mu\nu}$, defined as
$\Delta=\frac{1}{\sqrt{g}}\frac{\partial}{\partial x^{\mu}}\left(  \sqrt
{g}g^{\mu\nu}\frac{\partial}{\partial x^{\nu}}\right)  $, that is, $\Delta
\phi=g^{\mu\nu}\phi_{,\mu\nu}-\Gamma^{\mu}\phi_{\mu}$.$~$We remark that, when
$T_{\mu}^{\mu}$ corresponds to the cosmological constant, then the latter
equation of motion reduces to the usual Klein-Gordon equation for a modified
potential and that the chameleon mechanism is eliminated.

On the other hand, when $T^{\mu\nu}$ describes a pressureless ideal gas with
energy density $\rho_{m}$ the modified Klein-Gordon equation becomes \cite{w5}%
\begin{equation}
\Delta\phi=V_{,\phi}+f_{,\phi}\rho_{m}. \label{as1}%
\end{equation}
Last but not least, the matter source satisfies the equation of motion
$\nabla_{\nu}T^{eff~~\mu\nu}=0$, where $\nabla_{\nu}$ is the covariant
derivative defined by the Levi-Civita connection with respect to the metric
tensor $g_{\mu\nu}$. Consequently, from (\ref{as1}) it follows that when
$\rho_{m}$ is large enough the chameleon term has a non-zero contribution to
the mass of the scalar field.

For a spatially flat FLRW background geometry with scale factor $a\left(
t\right)  $ and line element
\begin{equation}
ds^{2}=-dt^{2}+a^{2}\left(  t\right)  \left(  dx^{2}+dy^{2}+dz^{2}\right)  ,
\end{equation}
the gravitational field equations are%

\begin{equation}
3H^{2}=\frac{1}{2}\dot{\phi}^{2}+V\left(  \phi\right)  -\rho_{m}f\left(
\phi\right)  ,\label{eq.01}%
\end{equation}%
\begin{equation}
2\dot{H}+3H^{2}=-\left(  \frac{1}{2}\dot{\phi}^{2}-V\left(  \phi\right)
-f\left(  \phi\right)  p_{m}\right)  ,\label{eq.02}%
\end{equation}
where $H=\frac{\dot{a}}{a}$ is the Hubble function; $\rho_{m}$ and $p_{m}$ are
the energy density and pressure components for a perfect fluid described by
the Lagrangian function $L_{m}\left(  \psi\left(  x^{\mu}\right)  \right)  $.
We have assumed the scalar field and the matter source to inherit the
symmetries of the background geometry which means that $\rho_{m}=\rho
_{m}\left(  t\right)  $ and $\phi=\phi\left(  t\right)  $.

The modified Klein-Gordon equation (\ref{as.0}) is expressed as
\begin{equation}
\ddot{\phi}+3H\dot{\phi}+V_{,\phi}+f_{,\phi}\rho_{m}=0,\label{eq.03}%
\end{equation}
while that for the perfect fluid the equation of motion is
\begin{equation}
\dot{\rho}_{m}+3H\left(  \rho_{m}+p_{m}\right)  -\dot{\phi}\left(  \ln
f\right)  _{,\phi}\rho_{m}=0\label{eq.04}%
\end{equation}

Let the perfect fluid to be an ideal gas with constant equation of state
parameter $w_{m}$, that is, $p_{m}=w_{m}\rho_{m}$. From equation (\ref{eq.04})
we find that \cite{lan1},
\begin{equation}
\rho_{m}=\rho_{m0}a^{-3\left(  w_{m}+1\right)  }f\left(  \phi\right)
\end{equation}
with $\rho_{m0}$ an integration constant.

By replacing in (\ref{ac.01}) we end with the point-like Lagrangian for the
field equations
\begin{equation}
\mathcal{L}\left(  a,\dot{a},\phi,\dot{\phi}\right)  =-3a\dot{a}^{2}+\frac
{1}{2}a^{3}\dot{\phi}^{2}-a^{3}V\left(  \phi\right)  -a^{-3w_{m}}F\left(
\phi\right)  ,~F\left(  \phi\right)  =f\left(  \phi\right)  ^{2}\,.\label{eq2}%
\end{equation}

Indeed, the field equations follow from the variation of $\mathcal{L}\left(
a,\dot{a},\phi,\dot{\phi}\right)  $ with respect to the dynamical variables
$a$ and $\phi$, while equation (\ref{eq.01}) can be seen as the Hamiltonian
constraint for the dynamical system. Indeed, Lagrangian (\ref{eq2}) is time
independent and it admits as Noether symmetry the vector field $\partial_{t}$,
with conservation law the Hamiltonian function, where from (\ref{eq.01}) it
follows that the \textquotedblleft energy\textquotedblright\ has a fixed value.

We continue our analysis with the investigation of the existence of functional
forms for the potential $V\left(  \phi\right)  $ and for the coupling
$F\left(  \phi\right)  $ where extra Noether symmetries exist. In particular,
we apply the theory of variational symmetries, that is, Noether's first
theorem, and we require the variation of the Action Integral (\ref{ac.01}) to
be invariant under one-parameter point transformations.

In the following we consider that $w_{m}\in\lbrack0,1)$. The limit $w_{m}=1$
corresponds to the Chiral model which has been studied before in \cite{ns2}.

\section{Variational symmetries}

\label{sec3}

In this Section we solve the classification problem for the admitted Noether
symmetries of the field equations with Lagrangian function (\ref{eq2}).\ 

\subsection{Symmetries of differential equations}

We review briefly the basic elements for the variational symmetries of
differential equations.

We assume the dynamical system of second-order differential equations
\begin{equation}
\mathbf{\ddot{x}}=\mathbf{\Omega}\left(  t,\mathbf{x},\mathbf{\dot{x}}\right)
\label{Lie.0}%
\end{equation}
which follows from the variation of the Lagrangian function%
\begin{equation}
\mathcal{L}=\mathcal{L}\left(  t,\mathbf{x},\mathbf{\dot{x}}\right)  .
\label{Lie.1}%
\end{equation}

In the augmented space $\{t,\mathbf{x}\}$ we consider the one-parameter point
transformation with generator the vector field
\begin{equation}
X=\xi\left(  t,\mathbf{x}\right)  \partial_{t}+\mathbf{\eta}\left(
t,\mathbf{x}\right)  \partial_{\mathbf{x}}.
\end{equation}
Then, according to Noether's first theorem, if there exist a boundary function
$g\left(  t,\mathbf{x},\mathbf{\dot{x}}\right)  $ such that the following
condition be true%
\begin{equation}
X^{\left[  1\right]  }\mathcal{L}+\mathcal{L}\dot{\xi}=\dot{g}, \label{Lie.5}%
\end{equation}
then the variation of the Action Integral with Lagrangian function
$\mathcal{L}\left(  t,\mathbf{x},\mathbf{\dot{x}}\right)  $ remains invariant
under the application of the one-parameter point transformation with generator
$X$.

When (\ref{Lie.5}) is true, the vector field $X$ is a variational symmetry, or
a Noether symmetry, for the dynamical system (\ref{Lie.0}). $X^{\left[
1\right]  }$ is the first prolongation \ of $X\,\ $in the jet space $\left\{
t,\mathbf{x},\mathbf{\dot{x}}\right\}  $ defined as%
\begin{equation}
X^{\left[  1\right]  }=X+\left(  \mathbf{\dot{\eta}}-\mathbf{\dot{x}}\dot{\xi
}\right)  \partial_{\mathbf{\dot{x}}}.
\end{equation}

Noether's second theorem states that for every variational symmetry vector
there exists a corresponding conservation law. Thus, if $X$ is a Noether
symmetry vector field, the function $I\left(  X\right)  $ defined as
\begin{equation}
I\left(  X\right)  =\xi\left(  \frac{\partial\mathcal{L}}{\partial
\mathbf{\dot{x}}}\dot{x}-\mathcal{L}\right)  -\frac{\partial\mathcal{L}%
}{\partial\mathbf{\dot{x}}}\mathbf{\eta}+g, \label{Lie.6}%
\end{equation}
is a conservation law for the dynamical system (\ref{Lie.0}).

\subsection{Variational symmetries in chameleon cosmology}

For the Lagrangian model (\ref{eq2}) with the chameleon mechanics, the
augmented space has dimension three and is $\left\{  t,a,\phi\right\}  $. We
assume the generator $X$ of a one-parameter point transformation in the base
manifold,
\begin{equation}
X=\xi\left(  t,a,\phi\right)  \partial_{t}+\eta_{a}\left(  t,a,\phi\right)
\partial_{a}+\eta_{\phi}\left(  t,a,\phi\right)  \partial_{\phi},
\label{Lie.3}%
\end{equation}
with first prolongation
\begin{equation}
X^{\left[  1\right]  }=X+\left(  \dot{\eta}_{a}-\dot{a}\dot{\xi}\right)
\partial_{\dot{a}}+\left(  \dot{\eta}_{\phi}-\dot{\phi}\dot{\xi}\right)
\partial_{\dot{\phi}}. \label{Lie.7}%
\end{equation}

From the symmetry condition (\ref{Lie.5}) and for Lagrangian function
(\ref{eq2}) we determine a linear system of first-order partial differential
equations with respect the coefficient functions $\xi,~\eta_{a}$ and
$\eta_{\phi}$ of the symmetry vector $X$. This system is presented in Appendix
\ref{appen1}. The solution of the symmetry condition (\ref{Lie.5}) depends
upon the functional form of the potential function $V\left(  \phi\right)  $
and of the coupling scalar $F\left(  \phi\right)  $. As we discussed above,
for arbitrary functions $V\left(  \phi\right)  $ and $F\left(  \phi\right)  $
the variational symmetry $X^{1}=\partial_{t}$ exists always with conservation
law the Hamiltonian $H=\frac{\partial\mathcal{L}}{\partial\mathbf{\dot{x}}%
}\dot{x}-L$. However, from (\ref{eq.01}) it follows $H=0$.

We omit the presentation of the calculations.\ We remark that in the following
we consider $F\left(  \phi\right)  $ is a non-constant function.

\subsubsection{Case A}

For the potential function%
\[
V_{A}\left(  \phi\right)  =\left(  V_{1}e^{\frac{\sqrt{6}}{4}\phi}%
-V_{2}e^{-\frac{\sqrt{6}}{4}\phi}\right)  ^{2}
\]
and the coupling function%
\begin{equation}
F_{A}\left(  \phi\right)  =F_{0}\left(  V_{1}e^{\frac{\sqrt{6}}{4}\phi}%
-V_{2}e^{-\frac{\sqrt{6}}{4}\phi}\right)  ^{-2w_{m}},
\end{equation}
i.e. $F_{A}\left(  \phi\right)  =F_{0}\left(  V_{A}\left(  \phi\right)
\right)  ^{-w_{m}}$, there exist the variational symmetries%
\begin{equation}
X^{2}=\frac{1}{\sqrt{a}}\left(  V_{1}e^{\frac{\sqrt{6}}{4}\phi}+V_{2}%
e^{-\frac{\sqrt{6}}{4}\phi}\right)  \partial_{a}-\frac{\sqrt{6}}{a^{\frac
{3}{2}}}\left(  V_{1}e^{\frac{\sqrt{6}}{4}\phi}+V_{2}e^{-\frac{\sqrt{6}}%
{4}\phi}\right)  \partial_{\phi}%
\end{equation}
and%
\begin{equation}
X^{3}=t\left(  \frac{1}{\sqrt{a}}\left(  V_{1}e^{\frac{\sqrt{6}}{4}\phi}%
+V_{2}e^{-\frac{\sqrt{6}}{4}\phi}\right)  \partial_{a}-\frac{\sqrt{6}%
}{a^{\frac{3}{2}}}\left(  V_{1}e^{\frac{\sqrt{6}}{4}\phi}+V_{2}e^{-\frac
{\sqrt{6}}{4}\phi}\right)  \partial_{\phi}\right)  .
\end{equation}
The corresponding boundary functions $g\left(  t,a,\phi\right)  $ are
$g\left(  X^{2}\right)  =const$ and $g\left(  X^{3}\right)  =-4a^{\frac{3}{2}%
}\left(  V_{1}e^{\frac{\sqrt{6}}{4}\phi}+V_{2}e^{-\frac{\sqrt{6}}{4}\phi
}\right)  $.

Consequently, from Noether's second theorem and expression (\ref{Lie.6}) the
conservation laws are
\begin{equation}
I\left(  X^{2}\right)  =-6\left(  V_{1}e^{\frac{\sqrt{6}}{4}\phi}%
+V_{2}e^{-\frac{\sqrt{6}}{4}\phi}\right)  a^{\frac{1}{2}}\dot{a}-\sqrt
{6}a^{\frac{3}{2}}\left(  V_{1}e^{\frac{\sqrt{6}}{4}\phi}+V_{2}e^{-\frac
{\sqrt{6}}{4}\phi}\right)  \dot{\phi}%
\end{equation}
and%
\begin{equation}
I\left(  X^{3}\right)  =tI\left(  X^{2}\right)  +4a^{\frac{3}{2}}\left(
V_{1}e^{\frac{\sqrt{6}}{4}\phi}+V_{2}e^{-\frac{\sqrt{6}}{4}\phi}\right)  .
\end{equation}

We remark that for $V_{1}V_{2}=0$ \ it follows that $V_{A}\left(  \phi\right)
\simeq e^{\pm\frac{\sqrt{6}}{2}\phi}$ and $F_{A}\left(  \phi\right)  \simeq
e^{\mp\frac{\sqrt{6}}{2}w_{m}\phi}$. Furthermore, for $V_{2}=V_{1}$, it
follows that $V_{A}\left(  \phi\right)  \simeq\sinh^{2}\left(  \frac{\sqrt{6}%
}{4}\phi\right)  $,~$F_{A}\left(  \phi\right)  \simeq\sinh^{-2w_{m}}\left(
\frac{\sqrt{6}}{4}\phi\right)  $. On the other hand for $V_{2}=-V_{1}$ we find
that $V_{A}\left(  \phi\right)  \simeq\cosh^{2}\left(  \frac{\sqrt{6}}{4}%
\phi\right)  $,~$F_{A}\left(  \phi\right)  \simeq\sinh^{-2w_{m}}\left(
\frac{\sqrt{6}}{4}\phi\right)  $.

\subsubsection{Case B}

The dynamical system with potential function
\begin{equation}
V_{B}\left(  \phi\right)  =\left(  V_{1}e^{\frac{\sqrt{6}}{4}\phi}%
-V_{2}e^{-\frac{\sqrt{6}}{4}\phi}\right)  ^{2}+\frac{4}{3}\Lambda
\end{equation}
and the coupling function $F_{B}\left(  \phi\right)  =F_{A}\left(
\phi\right)  $ the resulting Noether symmetries and boundary values are
\begin{equation}
Y^{1}=e^{\sqrt{\Lambda}t}X^{2}~,~g\left(  Y^{1}\right)  =-4\sqrt{\Lambda
}e^{\sqrt{\Lambda}t}a^{\frac{3}{2}}\left(  V_{1}e^{\frac{\sqrt{6}}{4}\phi
}+V_{2}e^{-\frac{\sqrt{6}}{4}\phi}\right)  ,
\end{equation}%
\begin{equation}
Y^{1}=e^{-\frac{2\sqrt{3\Lambda}}{3}t}X^{2}~,~g\left(  Y^{2}\right)
=4\sqrt{\Lambda}e^{-\sqrt{\Lambda}t}a^{\frac{3}{2}}\left(  V_{1}e^{\frac
{\sqrt{6}}{4}\phi}+V_{2}e^{-\frac{\sqrt{6}}{4}\phi}\right)  .
\end{equation}

The corresponding conservation laws are%
\begin{equation}
I\left(  Y^{1}\right)  =e^{\sqrt{\Lambda}t}\left(  I\left(  X^{2}\right)
-\sqrt{\Lambda}~g\left(  Y^{1}\right)  \right)
\end{equation}
and%
\begin{equation}
I\left(  Y^{2}\right)  =e^{-\sqrt{\Lambda}t}\left(  I\left(  X^{2}\right)
+\sqrt{\Lambda}g\left(  Y^{2}\right)  \right)  .
\end{equation}

Indeed, from these two time-dependent conservation laws we can construct the
time-independent function~$I\left(  Y^{1}Y^{2}\right)  =I\left(  Y^{1}\right)
I\left(  Y^{2}\right)  $, that is,%
\begin{equation}
I\left(  Y^{1}Y^{2}\right)  =\left(  I\left(  X^{2}\right)  \right)
^{2}-16\Lambda a^{2}\left(  V_{1}e^{\frac{\sqrt{6}}{4}\phi}+V_{2}%
e^{-\frac{\sqrt{6}}{4}\phi}\right)  ^{2}.
\end{equation}

For specific values of the free parameters $V_{1}$ and $V_{2}$ the scalar
field potential is expressed by specific functions. Indeed, for $V_{1}%
V_{2}=0\,$, we find the exponential potential $V_{B}\left(  \phi\right)
\simeq e^{\pm\frac{\sqrt{6}}{2}\phi}+\frac{4}{3}\Lambda$. For $V_{2}=V_{1}$,
the potential function is $V_{B}\left(  \phi\right)  \simeq\sinh^{2}\left(
\frac{\sqrt{6}}{4}\phi\right)  +\frac{4}{3}\Lambda$ while for $V_{2}=-V_{1}$.
It follows that $V_{B}\left(  \phi\right)  \simeq\cosh^{2}\left(  \frac
{\sqrt{6}}{4}\phi\right)  +\frac{4}{3}\Lambda$.

\subsubsection{Case C}

In the third case provide by the symmetry analysis, the scalar field potential
is expressed as%
\begin{equation}
V_{C}\left(  \phi\right)  =V_{0}e^{\lambda\phi}%
\end{equation}
with coupling function
\begin{equation}
F_{C}\left(  \phi\right)  =F_{0}e^{\frac{\phi}{2}\lambda\left(  w_{m}%
-1\right)  }.
\end{equation}
This case correspond to the\ WIS and it has been widely studied before in
\cite{ns1}.

The nontrivial variational symmetry is
\begin{equation}
Z^{1}=2t+\frac{2}{3}a\partial_{t}+\frac{4}{\lambda}\partial_{\phi}%
\end{equation}
with corresponding conservation law%
\begin{equation}
I\left(  Z^{1}\right)  =-4a^{2}\dot{a}+\frac{4}{\lambda}a^{3}\dot{\phi
}\text{.}%
\end{equation}

\subsubsection{Case D}

In the last case of the classification scheme the scalar field potential
vanishes, that is, $V_{D}\left(  \phi\right)  =0$ and $F\left(  \phi\right)
=F_{0}e^{-\frac{\sqrt{6}}{2}w_{m}\phi}$.

The Noether symmetry vectors are
\begin{equation}
Z^{2}=2t+\frac{2}{3}a\partial_{a}+\frac{2\sqrt{6}\left(  1-w_{m}\right)
}{w_{m}}\partial_{\phi},
\end{equation}
\begin{equation}
Z^{3}=e^{\frac{\sqrt{6}}{4}\phi}\left(  \frac{1}{\sqrt{a}}\partial_{a}%
-\frac{\sqrt{6}}{a^{\frac{3}{2}}}\partial_{\phi}\right)
\end{equation}
and
\begin{equation}
Z^{4}=te^{\frac{\sqrt{6}}{4}\phi}\left(  \frac{1}{\sqrt{a}}\partial_{a}%
-\frac{\sqrt{6}}{a^{\frac{3}{2}}}\partial_{\phi}\right)  .~g\left(
Z^{4}\right)  =-4a^{\frac{3}{2}}e^{\frac{\sqrt{6}}{4}\phi}.
\end{equation}

The corresponding conservation laws are derived to be%
\begin{equation}
I\left(  Z^{2}\right)  =-4a^{2}\dot{a}+\frac{2\sqrt{6}\left(  1-w_{m}\right)
}{w_{m}}a^{3}\dot{\phi},
\end{equation}%
\begin{equation}
I\left(  Z^{3}\right)  =e^{\frac{\sqrt{6}}{4}\phi}\left(  -6a^{\frac{1}{2}%
}\dot{a}-\sqrt{6}a^{\frac{3}{2}}\dot{\phi}\right)
\end{equation}
and%
\begin{equation}
I\left(  Z^{4}\right)  =tI\left(  Z^{3}\right)  -g\left(  Z^{4}\right)  \,.
\end{equation}

This integrable model is a special case of case $C$, for $V_{0}$ and
$\lambda=-\sqrt{6}\frac{w_{m}}{w_{m}-1}$.

We conclude that for these four sets of scalar fields and coupling functions
the field equations admit two conservation laws which are independent and in
involution. The number of independent conservation laws is the same as the
number of dynamical variables. Consequently for the above four sets of the
free functions the cosmological field equations form a Liouville integrable
dynamical system.

\section{Analytic solutions}

\label{sec4}

In this Section we continue our analysis with the derivation of new analytic
solutions for the cosmological models determined by the Noether symmetry
analysis presented above. The cosmological models of cases C and D have been
investigated in detail in \cite{ns1}. Hence, in this work we focus upon the
derivation of the solutions for the hyperbolic potential functions of cases A
and B. Furthermore, the evolution of physical parameters and the asymptotic
solutions of the cosmological solutions are discussed.

\subsection{Cases A \& B}

Cases A and B share the same coupling function $F\left(  \phi\right)  $ while
the scalar field potential functions differ in a constant parameter. Consider
now the scalar field potential of case $B$, that is,
\begin{equation}
V_{B}\left(  \phi\right)  =\left(  V_{1}e^{\frac{\sqrt{6}}{4}\phi}%
-V_{2}e^{-\frac{\sqrt{6}}{4}\phi}\right)  ^{2}+\frac{4}{3}\Lambda
\end{equation}
and the new new set of dependent variables $\left(  a,\phi\right)
\rightarrow\left(  x,y\right)  $ with transformation rule%
\[
a=\left(  x^{2}-y^{2}\right)  ^{\frac{1}{3}}~,~\phi=\frac{2\sqrt{6}}{3}\arctan
h\left(  \frac{y}{x}\right)  .
\]

In the new variables the cosmological point-like Lagrangian (\ref{eq2})
becomes%
\begin{equation}
\mathcal{L}_{A}\left(  x,\dot{x},y,\dot{y}\right)  =\frac{4}{3}\left(  \dot
{x}^{2}-\dot{y}^{2}\right)  +\left(  V_{1-2}x+V_{1+2}y\right)  ^{2}+\frac
{4}{3}\Lambda\left(  x^{2}-y^{2}\right)  +F_{0}\rho_{m0}\left(  V_{1-2}%
x+V_{1+2}y\right)  ^{-2w_{m}},
\end{equation}
where in the new variables the field equations are%
\begin{equation}
\frac{4}{3}\left(  \dot{x}^{2}-\dot{y}^{2}\right)  -\left(  V_{1-2}%
x+V_{1+2}y\right)  ^{2}-\frac{4}{3}\Lambda\left(  x^{2}-y^{2}\right)
-F_{0}\rho_{m0}\left(  V_{1-2}x+V_{1+2}y\right)  ^{-2w_{m}}=0,
\end{equation}%
\begin{equation}
\ddot{x}-\frac{3}{4}V_{1-2}\left(  \left(  V_{1-2}x+V_{1+2}y\right)
+w_{m}F_{0}\rho_{m0}\left(  V_{1-2}x+V_{1+2}y\right)  ^{-2w_{m}-1}\right)
-\Lambda x=0,
\end{equation}
and
\begin{equation}
\ddot{y}+\frac{3}{4}V_{1+2}\left(  \left(  V_{1-2}x+V_{1+2}y\right)
+w_{m}F_{0}\rho_{m0}\left(  V_{1-2}x+V_{1+2}y\right)  ^{-2w_{m}-1}\right)
-\Lambda x=0
\end{equation}
with $V_{1-2}=V_{1}-V_{2}$ and $V_{1+2}=V_{1}+V_{2}$.

We proceed by considering some specific values of the free parameters $V_{1}%
$,~$V_{2}$. In particular we investigate the cases for which (a) $V_{1}%
+V_{2}=0$, (b) $V_{1}-V_{2}=0$ and (c) $V_{1}$,~$V_{2}$ are arbitrary.

\subsubsection{Solution for $V_{1}+V_{2}=0$}

For $V_{1}+V_{2}=0$ the field equations are simplified to the following
expressions%
\begin{equation}
\frac{4}{3}\left(  \dot{x}^{2}-\dot{y}^{2}\right)  -V_{1-2}x^{2}-\frac{4}%
{3}\Lambda\left(  x^{2}-y^{2}\right)  -F_{0}\rho_{m0}\left(  V_{1-2}x\right)
^{-2w_{m}}-=0, \label{sd0}%
\end{equation}%
\begin{equation}
\ddot{x}-\frac{3}{4}\left(  V_{1-2}\right)  \left(  V_{1-2}x+w_{m}F_{0}%
\rho_{m0}\left(  V_{1-2}x\right)  ^{-2w_{m}-1}\right)  -\Lambda x=0
\label{sd1}%
\end{equation}
and
\begin{equation}
\ddot{y}-\Lambda x=0. \label{sd3}%
\end{equation}

For $\Lambda=0$ we observe that the Noetherian conservation laws $I\left(
X^{2}\right)  $ and $I\left(  X^{3}\right)  $ are $I\left(  X^{2}\right)
\simeq\dot{y}$ and $I\left(  X^{3}\right)  \simeq t\dot{y}-y$. On the other
hand, for $\Lambda\neq0$, the conservation laws $I\left(  Y^{1}\right)  $ and
$I\left(  Y^{2}\right)  $ are $I\left(  Y^{1}\right)  =e^{\sqrt{\Lambda}%
t}\left(  \dot{y}-\sqrt{\Lambda}y\right)  $ and $I\left(  Y^{2}\right)
=e^{-\sqrt{\Lambda}t}\left(  \dot{y}+\sqrt{\Lambda}y\right)  $ from which we
can defined the time-independent conservation law~$I\left(  Y^{1}Y^{2}\right)
=\dot{y}^{2}-y^{2}$.

The solution of equation (\ref{sd3}) is $y\left(  t\right)  \simeq\left(
t-t_{0}\right)  $ for $\Lambda=0$ and $y\left(  t\right)  \simeq\sqrt
{\frac{I\left(  Y^{1}Y^{2}\right)  }{\Lambda}}\sinh\left(  \sqrt{\Lambda
}\left(  t-t_{0}\right)  \right)  $ when $\Lambda\neq0$.

With the use of the conservation laws, the constraint equation (\ref{sd0})
becomes%
\begin{equation}
\frac{dx}{\sqrt{\frac{3}{4}\left(  I_{0}\right)  ^{2}+\frac{3}{4}\left(
\left(  \left(  V_{1-2}\right)  ^{2}+\frac{16}{9}\Lambda\right)  x^{2}%
+F_{0}\rho_{m0}\left(  V_{1-2}x\right)  ^{-2w_{m}}\right)  }}=\pm dt\text{ }
\label{sd4}%
\end{equation}
with $I_{0}=I\left(  X^{2}\right)  $ $\ $or $I\left(  X^{2}\right)  $
$=I\left(  Y^{1}Y^{2}\right)  $. Thus the analytic solution is expressed in
terms of the elliptic integral (\ref{sd4}). We observe that, when the
chameleon mechanism is eliminated, the analytic solution for the unified dark
matter potential for the quintessence field \cite{ns11} is recovered.

Indeed, for large values of $x$, because $w_{m}\geq0$, the surviving terms in
(\ref{sd1}) are
\begin{equation}
\ddot{x}-\frac{3}{4}\left(  \left(  V_{1-2}x\right)  ^{2}+\frac{16}{9}%
\Lambda\right)  x\simeq0.
\end{equation}

The chameleon mechanism does not contribute to the field equations and the
model is reduced to that of quintessence theory. The asymptotic solution at
this limit is expressed by the closed-form function
\begin{equation}
x\left(  t\right)  \simeq x_{1}\sinh\left(  \sqrt{\frac{3}{4}\left(  \left(
V_{1-2}x\right)  ^{2}-\frac{16}{9}\Lambda\right)  }t+x_{2}\right)  .
\end{equation}

Consider now that $\frac{3}{4}\left(  \left(  V_{1-2}x\right)  ^{2}-\frac
{16}{9}\Lambda\right)  >0$, then the scale factor for \thinspace$x>>1$, that
is, for $t>>0$, it is asymptotically described by the hyperbolic function
\begin{equation}
a\left(  t\right)  \simeq\left(  x_{1}\right)  ^{\frac{2}{3}}\sinh^{\frac
{2}{3}}\left(  \sqrt{\frac{3}{4}\left(  \left(  V_{1-2}x\right)  ^{2}%
-\frac{16}{9}\Lambda\right)  }t+x_{2}\right)  . \label{sd.4a}%
\end{equation}
The latter describes the $\Lambda$CDM universe \cite{ns11}. Recall that for
the $\Lambda$CDM model the scale factor is $a\left(  t\right)  =\left(
\frac{\Omega_{m}}{\Omega_{\Lambda}}\right)  ^{\frac{1}{3}}\sinh^{\frac{2}{3}%
}\left(  \frac{3}{2}H_{0}\sqrt{\Omega_{\Lambda}}t\right)  $.

\begin{figure}[ptb]
\centering\includegraphics[width=1\textwidth]{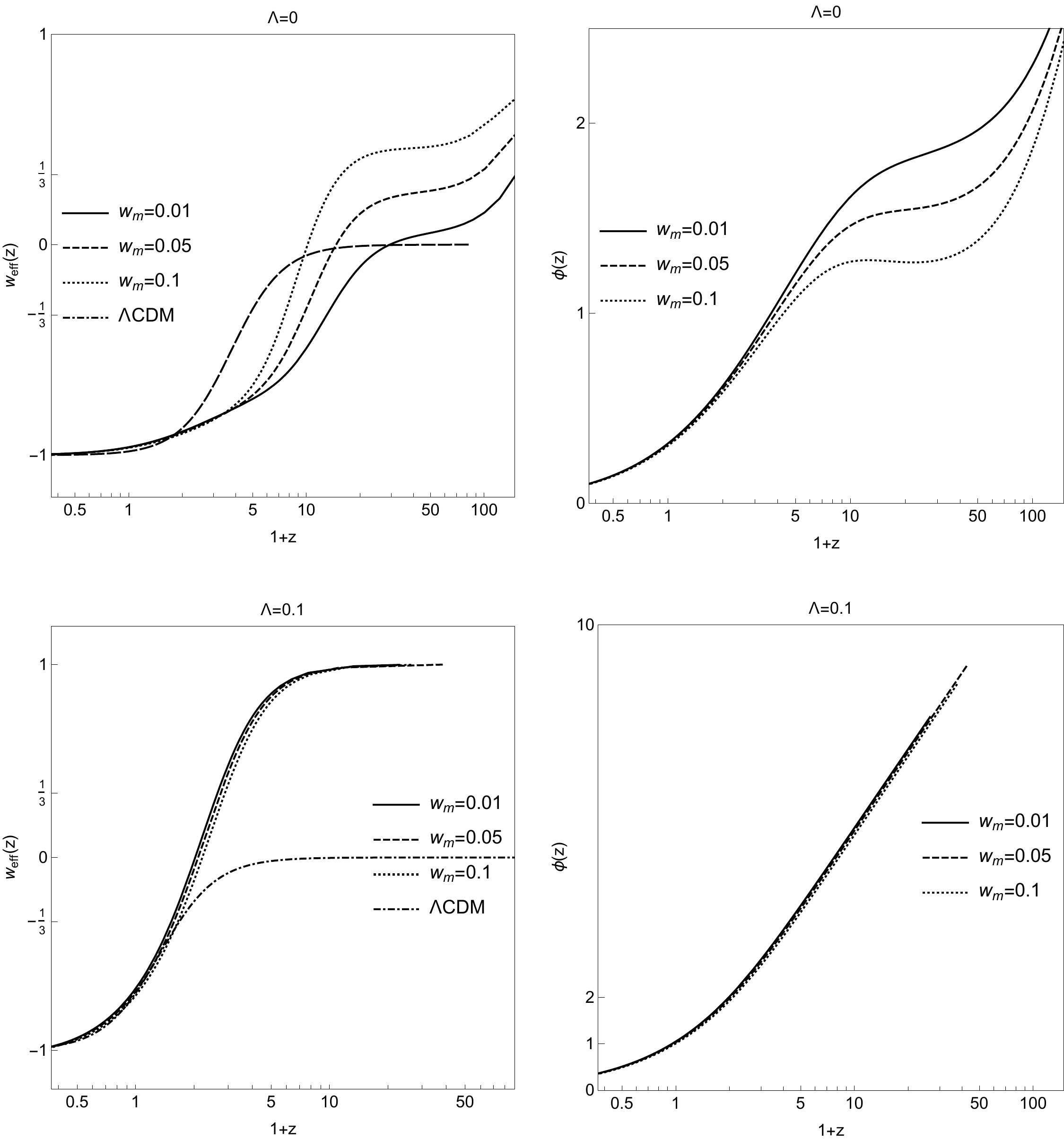}\caption{Qualitative
evolution of the effective equation of state parameter $w_{eff}\left(
z\right)  $ (left column) and of the scalar field $\phi\left(  z\right)  $
(right column) for the scalar field potential of cases A and B for
$V_{1}+V_{2}=0$. The plots at the first row are for $\Lambda=0$ and
$y=I_{0}t.$ In the second row the plots are for $\Lambda=0.1$ and $y\left(
t\right)  =\sqrt{\frac{I_{0}}{\Lambda}}\sinh\left(  \sqrt{\Lambda}t\right)  .$
\ For the plots we selected $\left(  V_{1},I_{0},F_{0}\rho_{m0}\right)
=\left(  1,0.1,0.01\right)  $ and initial condition $x\left(  0\right)  $ such
that $a\left(  0\right)  \simeq0$. Solid lines are for $w_{m}=0.01$, dashed
lines are for $w_{m}=0.05$ and dotted lines are for $w_{m}=0.1$. Dashed-dotted
lines are for the $\Lambda$CDM solution with $a\left(  t\right)  =a_{0}%
\sinh^{\frac{2}{3}}\left(  \omega t\right)  $ and $w_{eff}=-\tanh^{2}\left(
\omega t\right)  $. We observe that in the late universe the model mimics the
$\Lambda$CDM theory with de Sitter universe as a future attractor.}%
\label{fig1}%
\end{figure}

In Fig. \ref{fig1} we present the qualitative evolution for the effective
equation of state parameter $w_{eff}=-1-\frac{2}{3}\frac{\dot{H}}{H^{2}}$ and
for the scalar field $\phi$ for various values of the free parameters. We
observe that in the late universe; that is, for small values of the redshift,
the solution is very close to that of $\Lambda$CDM cosmology with the de
Sitter solution as a future attractor.

\subsubsection{Solution for $V_{1}-V_{2}=0$}

We assume now the case for which $V_{1}-V_{2}=0$. In this case the field
equations are
\begin{equation}
\frac{4}{3}\left(  \dot{x}^{2}-\dot{y}^{2}\right)  -\left(  V_{1+2}y\right)
^{2}-F_{0}\rho_{m0}\left(  V_{1+2}y\right)  ^{-2w_{m}}=0,
\end{equation}%
\begin{equation}
\ddot{x}-\Lambda y=0
\end{equation}
and
\begin{equation}
\ddot{y}+\frac{3}{4}V_{1+2}\left(  V_{1+2}y+w_{m}F_{0}\rho_{m0}\left(
V_{1+2}y\right)  ^{-2w_{m}-1}\right)  -\Lambda y=0,
\end{equation}
where now the corresponding Noetherian conservation laws are expressed as
$I\left(  X^{2}\right)  \simeq\dot{x}$ and $I\left(  X^{3}\right)  \simeq
t\dot{x}-x$, for $\Lambda=0$; or $I\left(  Y^{1}\right)  =e^{\sqrt{\Lambda}%
t}\left(  \dot{x}-\sqrt{\Lambda}X\right)  $ and $I\left(  Y^{2}\right)
=e^{-\sqrt{\Lambda}t}\left(  \dot{x}+\sqrt{\Lambda}x\right)  $ when
$\Lambda\neq0$

This dynamical system has the same solution process as in the previous case
with $V_{1}+V_{2}=0$. We omit the analysis of this case.

\subsubsection{Solution for arbitrary $V_{1},~V_{2}$}

We define the new variable $\,Z=V_{1+2}y-V_{1-2}x$ and we write the field
equations as%
\begin{align}
0  &  =4\left(  2V_{1}\dot{y}-\dot{Z}\right)  \left(  2V_{2}\dot{y}-\dot
{Z}\right)  -F_{0}\rho_{m0}Z^{-2w_{m}}+\nonumber\\
&  -\frac{1}{V_{1-2}^{2}}\left(  16V_{1}V_{2}\Lambda y^{2}-8\left(
V_{1+2}\right)  \Lambda y+\left(  3\left(  V_{1-2}^{2}+4\Lambda\right)
Z^{2}\right)  \right)  ~,
\end{align}%
\begin{equation}
0=\ddot{Z}+\left(  3V_{1}V_{2}-\Lambda\right)  Z-V_{1}V_{2}F_{0}w_{m}\rho
_{m0}Z^{-1-2w_{m}} \label{ars1}%
\end{equation}
and
\begin{equation}
0=\ddot{y}-\Lambda y+\frac{3}{4}\left(  V_{1+2}\right)  F_{0}w_{m}\rho
_{m0}Z^{-1-2w_{m}}-\frac{3}{4}\left(  V_{1+2}\right)  Z. \label{ars2}%
\end{equation}

We remark that equation (\ref{ars1}) is of the form of equation (\ref{sd1}).
Thus the analytic solution is expressed in terms of elliptic integrals. By
replacing into (\ref{ars2}) we find the second-order differential equation of
the form $\ddot{y}-\Lambda y=h\left(  t\right)  $, which is an integrable equation.

Consider now the simple Subcase with $\Lambda_{1}\Lambda_{2}=0$, say
$\Lambda_{2}=0$. \ Then the field equations become
\begin{equation}
-4\left(  2V_{1}\dot{y}-\dot{Z}\right)  \dot{Z}-\frac{1}{V_{1}^{2}}\left(
-8\left(  V_{1}\right)  \Lambda y+\left(  3\left(  V_{1}^{2}+4\Lambda\right)
Z^{2}\right)  \right)  -F_{0}\rho_{m0}Z^{-2w_{m}}=0,
\end{equation}%
\begin{equation}
\ddot{Z}-\Lambda Z=0
\end{equation}
and
\begin{equation}
\ddot{y}-\Lambda y+\frac{3}{4}\left(  V_{1}\right)  F_{0}w_{m}\rho
_{m0}Z^{-1-2w_{m}}-\frac{3}{4}\left(  V_{1}\right)  Z=0.
\end{equation}
For $\Lambda=0$ the analytic solution of this system is
\begin{equation}
Z=Z_{1}t+Z_{2},
\end{equation}%
\begin{equation}
y\left(  t\right)  =\frac{1}{8}\left(  \frac{3F_{0}\rho_{m0}}{Z_{1}^{2}\left(
2w_{m}-1\right)  }\left(  Z_{1}t+Z_{2}\right)  ^{1-2w_{m}}-\left(
Z_{1}t+Z_{2}\right)  V_{1}t\right)  +y_{1}t+y_{2},
\end{equation}
with constraint equation $Z_{2}^{2}-\frac{4}{3V_{1}^{2}}Z_{1}\left(
Z_{1}-2V_{1}y_{1}\right)  =0$. For $\Lambda\neq0$ the closed-form solution is
expressed in terms of the hyperbolic function. Thus we omit it. However, it is
easy that for $\Lambda\neq0$ the de Sitter universe is the late-time attractor
of this cosmological model.

\begin{figure}[ptb]
\centering\includegraphics[width=1\textwidth]{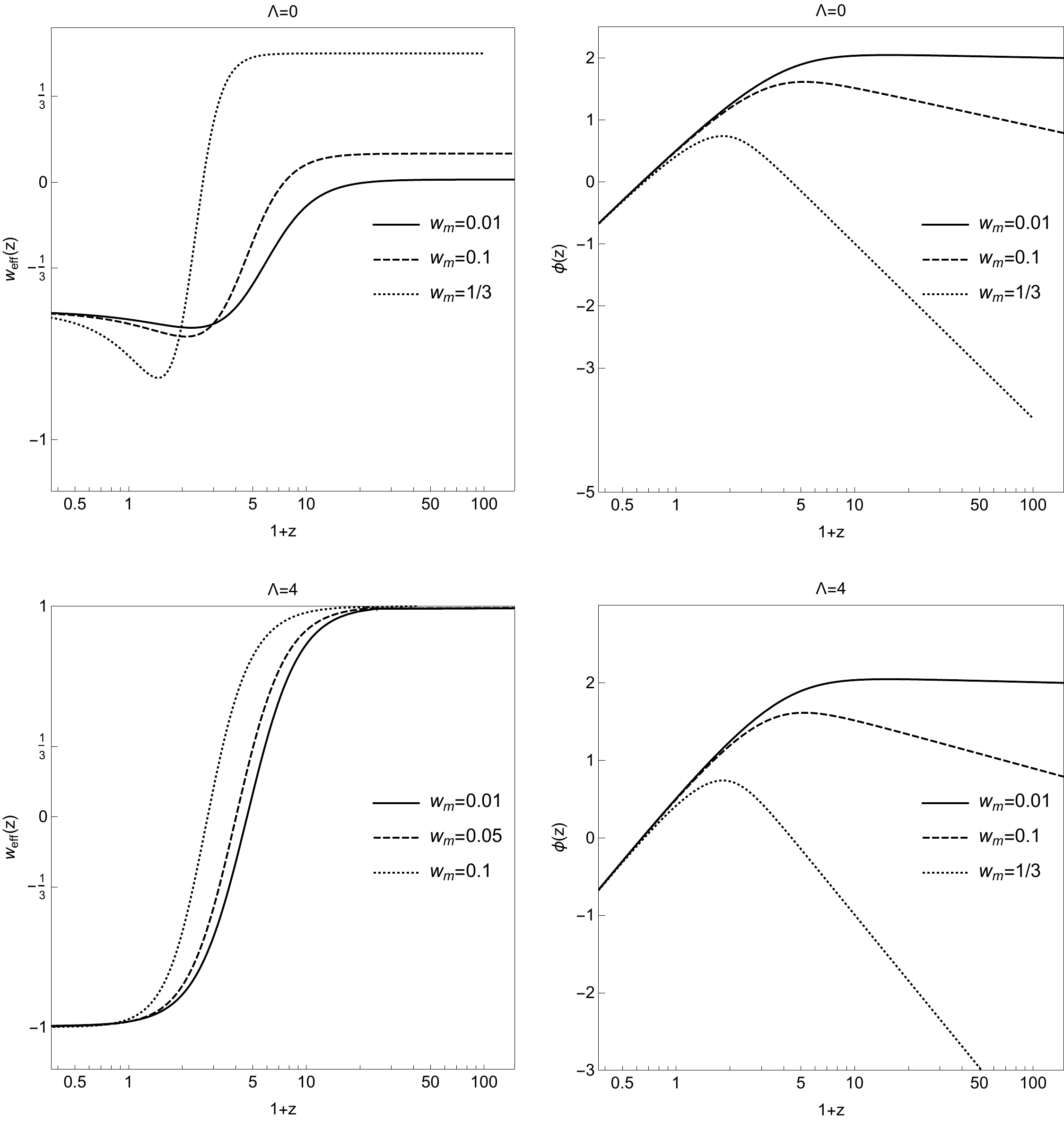}\caption{Qualitative
evolution of the effective equation of state parameter $w_{eff}\left(
z\right)  $ (left column) and of the scalar field $\phi\left(  z\right)  $
(right column) for the scalar field potential of cases A and B for $V_{2}=0$.
The plots in the first row are for $\Lambda=0$, $\left(  Z_{1},Z_{2}%
,V_{1},F_{0}\rho_{m0}\right)  =\left(  1,0,1,0.1\right)  $. In the second row
the plots are for $\Lambda=4$ and for $Z\left(  t\right)  =Z_{1}%
e^{\sqrt{\Lambda}t}$. Solid lines are for $w_{m}=0.01$, dashed lines are for
$w_{m}=0.1$ and dotted lines are for $w_{m}=\frac{1}{3}$. We observe that in
the late universe the late time attractor gives $w_{eff}<-\frac{1}{3}$ which
corresponds to an accelerated universe. For $\Lambda\neq0$ the future
attractor is the de Sitter universe. }%
\label{fig2}%
\end{figure}

In Fig. \ref{fig2} we present the qualitative evolution for the equation of
state parameter and the scalar field for this cosmological model for various
values of the free parameters.

\section{Conclusions}

\label{sec5}

Noether symmetry analysis is a powerful approach for the investigation of
integrable models in gravitational theories. Specifically, Noether's first
theorem is applied to constrain the unknown functions and parameters of the
gravitational Lagrangian such that nontrivial variational symmetries exist.
The latter are directly related with the existence of conservation laws for
the equations of motion. The exact relation between the variational symmetries
and the conservation laws is given by Noether's second theorem. In
gravitational physics the field equations are nonlinear and with free
functions and parameters. In \cite{ns11} it has been shown that the symmetry
analysis is a geometric selection rule for the determination of the unknown
functions and parameters. Indeed, the variational symmetries of the
gravitational Lagrangian are constructed by isometries and the homothecy of
the minisuperspace. This describes the geometry in the manifold on which the
dynamical variables of the gravitational theory lie.

In this study we considered the Noether symmetry analysis in chameleon
cosmology. We determined the scalar field potential and the coupling function
of the chameleon mechanism by the requirement that the gravitational field
equations admit Noether point symmetries. We derived four families of scalar
field potentials and coupling functions for which the field equations admit
conservation laws constructed from the second theorem of Noether. The equation
of state parameter for the ideal gas plays an important role in these
functions. Two of the derived models belong to the WIS family of models and
they have been studied before. However, for the two new models the
conservation laws indicate that the models are Liouville integrable and we
determined the analytic solutions in closed-form expressions.

From the analysis of the cosmological parameters and the asymptotic solution
(\ref{sd.4a}) we infer that for small values of the redshift; i.e. in the late
universe, these models can reproduce the $\Lambda$CDM with the de Sitter
universe to be a future attractor. That is a characteristic of special
interest for this theory because we have that the $\Lambda$CDM behaviour is
recovered without us to have required the existence of a pressureless fluid in
the cosmological model.

We conclude that the application of Noether symmetries in this model provided
us not only with analytic solutions which can be used as toy models, but also
as unified dark energy model. In future work we plan to investigate further if
this model can explain the cosmological observations.

%\textbf{Data Availability Statements:} Data sharing not applicable to this article as no datasets were generated or analyzed during the current study.

\begin{acknowledgments}
This work was partially financially supported in part by the National Research
Foundation of South Africa (Grant Numbers 131604). The author thanks the
support of Vicerrector\'{\i}a de Investigaci\'{o}n y Desarrollo
Tecnol\'{o}gico (Vridt) at Universidad Cat\'{o}lica del Norte through
N\'{u}cleo de Investigaci\'{o}n Geometr\'{\i}a Diferencial y Aplicaciones,
Resoluci\'{o}n Vridt No - 096/2022.
\end{acknowledgments}

%

%TCIMACRO{\TeXButton{Appendix}{\appendix}}%
%BeginExpansion
\appendix
%EndExpansion

\section{Noether symmetry conditions}

\label{appen1}

We present the determining system provided by the Noether symmetry condition
(\ref{Lie.5}) for the Lagrangian function (\ref{eq2}).%

\begin{equation}
\xi_{,a}=0~,~\xi_{,\phi}=0~,
\end{equation}%
\begin{equation}
6a\eta_{,t}^{a}+g_{,a}=0~,~a^{3}\eta_{,t}^{\phi}-g_{,\phi}=0,
\end{equation}%
\begin{equation}
2a\eta_{,a}^{a}+6\eta^{a}-a\xi_{,t}=0,
\end{equation}%
\begin{equation}
a^{2}\eta_{,a}^{\phi}-6\eta_{,\phi}^{a}=0,
\end{equation}%
\begin{equation}
3\eta^{a}+2a\eta_{,\phi}^{\phi}-\xi_{,a}=0,
\end{equation}%
\begin{equation}
3a^{2}\eta^{a}\left(  V+w_{m}\rho_{m0}a^{3w_{m}-3}F\right)  +a^{3}\eta^{\phi
}\left(  V_{,\phi}-\rho_{m0}a^{3w_{m}-3}F_{,\phi}\right)  +\left(
a^{3}V\left(  \phi\right)  -\rho_{m0}a^{-3w_{m}}F\right)  \xi_{,t}=-g_{,t}.
\end{equation}
The Noether symmetry conditions gives the coefficient components of the
Noether vector field $X$. The solution of the Noether conditions it depends on
the functional forms of the free functions $V\left(  \phi\right)  $ and
$F\left(  \phi\right)  $. Hence for various values of these functions as
presented in Section \ref{sec3} the admitted Noether symmetries differ.

\end{document}